\documentclass{JHEP3}

\usepackage{amsmath}
\input epsf
\usepackage{epsfig}
\usepackage{amssymb}
\usepackage{graphics}
\usepackage[active]{srcltx}
\usepackage{graphicx}
\usepackage{subfig}

\newcommand{\bea}{\begin{eqnarray}}
\newcommand{\eea}{\end{eqnarray}}
\newcommand{\bean}{\begin{eqnarray*}}
\newcommand{\eean}{\end{eqnarray*}}
\newcommand{\nn}{\nonumber \\}

\def\W #1{\widetilde{#1}}
\def\WH #1{\widehat{#1}}

\def\braket#1{\left\langle #1 \right\rangle}

\def\gb #1{ \left\langle #1 \right]}

\def\eref#1{(\ref{#1})}

\def\eps{\epsilon}

\def\vev{\braket}

\def\Spaa{\vev}

\def\Spab{\gb}

\newcommand{\cA}{{\cal A}}

\newcommand{\cI}{{\cal I}}

\def\Label#1{\label{#1}%
  \smash{\hbox to0pt{\raise1ex\hbox{\tiny[#1]}\hss}}}

\def\oneloop{\tiny\mbox{1-loop}}

\def\cyclic#1{\mbox{Cyclic}\{#1\}}

\def\spaa #1{\langle #1\rangle}
\def\spbb #1{[#1]}
\def\spab #1{\langle #1]}
\def\spba #1{[#1 \rangle}

\title{General expressions for extra-dimensional tree amplitudes and all-plus 1-loop integrands in $\mathcal{Q}$-cut representation}
\author{Yang An$^{\ddagger}$, Yi Li$^{\ddagger}$~~~~\\
$^\ddagger$ Zhejiang Institute of Modern Physics, Department of Physics,
 Zhejiang University, Hangzhou, 310027, P.R. China\\
 \\
E-mail: \email{anyangpeacefulocean@zju.edu.cn}, \email{liyiphysics@zju.edu.cn}}

\abstract{
In this paper, we give the general expressions for a special series of tree amplitudes of the Yang-Mills theory. This series of amplitudes have two adjacent massless spin-1 particles with extra-dimensional momenta and any number of positive helicity gluons. With special helicity choices, we use the spinor helicity formalism to express these n-point amplitudes in compact forms, and find a clever way to use the BCFW recursion relations to prove the results. Then these amplitudes are used to form the complete 1-loop all-plus integrand with any number of gluons, expressed in the $\mathcal{Q}$-cut representation.
}

\begin{document}

\section{Introduction}
Recently the $\mathcal{Q}$-cut construction \cite{Baadsgaard:2015twa} \cite{Huang:2015cwh} has been developed to compute complete loop integrands of massless field theory. When using $\mathcal{Q}$-cut representation to calculate loop integrands, a special series of on-shell tree amplitudes in general $D$-dimension are required. These tree amplitudes have two adjacent legs with extra-dimensional momenta and other legs with 4-dimensional momenta.

For all-plus 1-loop gluon amplitudes in Yang-Mills theory, the helicity of 4-dimensional particles (gluons) of the required tree amplitudes are all plus, which is just similar to the MHV amplitudes. Our motivation for this paper is to see if we can generalize these tree amplitudes to include any number of positive helicity gluons and find compact expressions for them. With these tree amplitudes we can calculate general expressions for the integrands of n-point 1-loop all-plus amplitudes in $\mathcal{Q}$-cut representation.

However, it is difficult to do this work by directly calculating a tremendous number of Feynman diagrams involving n gluons. Fortunately, calculations of multi-particle amplitudes have greatly developed in the last ten years, starting from the twistor string description of ${\cal N}=4$ Yang-Mills proposed 
by Witten~\cite{Witten:2003nn}.  New ideas have led to the development
of new powerful formalisms. The MHV vertex expansion (CSW)\cite{Cachazo:2004kj} and the BCFW recursion relations \cite{Britto:2004ap,Britto:2005fq} are two most important ones. Recently CHY formula \cite{Cachazo:2013hca}\cite{Cachazo:2013iea} is developed, which can be used to calculate scattering amplitudes in arbitrary dimension.

In this paper, to deal with our extra-dimensional cases, we focus on the BCFW recursion relations, which is an on-shell recursive method developed by Britto, Cachazo, Feng, and Witten, where an higher-point tree amplitude can be given by
the sum of products of a propagator and two on-shell lower-point amplitudes with shifted, complex momenta.  
Later, through the work of Badger, Glover, Khoze, and Svr\v cek
\cite{Badger:2005zh}, the BCFW recursion relations have been generalized to include massive particles with
spin. 

The BCFW recursion relations for massive particles are also applicable in our cases because each helicity state of a massless $D$-dimensional spin-1 particle is equivalent to a massive particle state. A massless spin-1 particle with $D$-dimensional momentum has $D-2$ polarization degrees of freedom instead of 2 for a 4-dimensional massless spin-1 particle. So we need to choose more helicity states and this series of extra-dimensional amplitudes contain more kinds of amplitudes. After choosing totally $D-2$ physical polarization vectors labeled by helicity $+$ $-$ and $S_a$($a\in \{4,5, \dots, D-1\}$), we can follow the similar treatments using the BCFW recursion relations in amplitudes with massive legs and apply the generalized spinor helicity formalism for massive amplitudes to express these tree amplitudes involving 2 adjacent extra-dimensional legs. Our results for this series of different amplitudes are concise and have a common structure astonishingly. 


In our cases, the two $D$-dimensional spin-1 particles are equivalent to two massive particles with equal mass and 4-dimensional momenta. We follow the generalization of spinor helicity formalism for massive particles to express our results in helicity amplitudes. By introducing a null reference vector $q_i$, a massive momentum $p_i$ can be decomposed along two lightlike directions. In this paper we set all $q_i=q$, so $p_i = p^{\bot}_i - \frac{m_i^2}{2 q\cdot p_i}\, q$. The amplitudes are then q-dependent. For an introduction to the approaches to generalize massless spinor helicity formalism, see \cite{Dittmaier:1998nn}. In \cite{Boels:2008du,Boels:2009bv,Boels:2010mj}, applications of BCFW recursion relations with massive particles can be found. Compact expressions for several towers of tree-level amplitudes with a complex scalar-antiscalar pair or a massive W-boson pair and any numbers of positive helicity gluons are given in \cite{Ferrario:2006np,Craig:2011ws,Kiermaier:2011cr}. Besides, in \cite{Craig:2011ws}, superamplitudes on the Coulomb-branch of N = 4 SYM are calculated from massive amplitudes, which implies the massive version of our extra-dimensional amplitudes can be used to study superamplitudes on the Coulomb-branch. 

To compare with treatments involving massive particles, we review a recent work, a four-dimensional formulation (FDF) of the extra-dimensional regularization of 1-loop scattering amplitudes \cite{Fazio:2014xea}.


This paper is organized as follows. In section \ref{notation}, we set up conventions, introduce polarization vectors with helicity choices + - $S_a$, compare with the formulation in \cite{Fazio:2014xea} and give compact expressions for these tree amplitudes. In section \ref{proof}, a proof by mathematical induction of the results using the BCFW recursion relations is given. Finally in section \ref{qcut}, we use the expressions to give general expressions for the compact integrands of all-plus 1-loop amplitudes, written in $\mathcal{Q}$-cut representation.

\section{Conventions and main results}
\label{notation}

In this section, we will first introduce our conventions 
and helicity choices, compared with four dimensional treatment as massive particles.
Then, to set our starting point, we give expressions for related 3-point tree amplitudes and 4-point tree amplitudes. Finally we summarize our main results for n-point tree amplitudes and analyze their structures.

We only need to calculate some of amplitudes, since color-ordered amplitudes have the cyclic property $\cA_n(1,2,\dots,n)=\cA_n(2,\dots,n,1)$, and the reflection property:  $\cA_n(1,2,\dots,n)=(-1)^n\cA_n(n,\dots,2,1)$.

\subsection{Helicity choices and polarization vectors}
\label{helicity}

The convention of $D$-dimensional metric is $\eta^{\mu\nu}=\mbox{diag}(+,-,-,\cdots,-)$ throughout the paper.
Let's denote an extra-dimensional momentum as
\bea \WH \ell=
(\ell,\vec{\mu})~~~,~~~\vec{\mu}=(\mu_1,\ldots,\mu_d)~,~~~\label{D-ell}\eea
where $\ell$ is the 4-dimensional component and $\vec{\mu}$ is a
vector in the extra $d$-dimension ($d=D-4$). For the Euclidean
$d$-dimensional space, the extra-dimensional basis $e_a$ for $a=1,...,d$ can be choosen as $\vec{\mu}_{e_a}=(\delta_{1a},\ldots,\delta_{ia},\ldots,\delta_{da})$ .  The massless
condition of $\WH \ell$ is then $\ell^2-\vec{\mu}^2=0$. 


By introducing a null auxiliary momentum q in 4-dimension and $q\cdot \ell\neq 0$, we can define the null momentum as
\bea \ell^{\bot}=\ell-{\vec{\mu}^2\over
\Spab{q|\ell|q}}q~~~,~~~~({\ell^{\bot}})^2=0~,~~~\label{ell-bot}\eea

The transverse condition $\eps\cdot \WH \ell=0$
tells that the transverse space is $(D-2)$-dimensional. 
By imposing the extra condition $\eps_i\cdot q=0$, the transverse polarization vectors are fixed. 
The polarization vectors in \cite{Huang:2015cwh} are listed here with $D-2$ helicity choices $+~-$ and $S_a$ for spin-1 particles
\bea &&\epsilon^+_\nu(\widehat{\ell},q)=
\Big({\spab{q|\gamma_\nu|\ell^{\bot}}\over
\sqrt{2}\spaa{q~\ell^{\bot}}},\vec{0}_d\Big)~~~,~~~\epsilon^-_\nu(\widehat{\ell},q)=
\Big({\spab{\ell^{\bot}|\gamma_\nu|q}\over
\sqrt{2}\spbb{\ell^{\bot}~q}},\vec{0}_d\Big)~,~~~\nonumber\\
&&\epsilon^{S_a}(\widehat{\ell},q)=\Big({2 \mu_a\over
\spab{q|\ell|q}}q,e_a\Big)~~~,~~~a=1,...,d~,~~~\label{Pol-D-1}\eea
where $e_a$ is
the basis vector of the extra $d$-dimension, $\vec{0}_d$ denotes the $d$-dimensional vanishing vector, and $\mu_a$ is the $a$-th
component of $\vec{\mu}$ in \eref{D-ell}. 
The rest two
polarization vectors are longitudinal and time-like which have no physical effect, defined as
\bea \epsilon^{L}(\widehat{\ell},q)=\WH
\ell~~~,~~~\epsilon^{T}(\widehat{\ell},q)=(q,\vec{0}_d)~.~~~\label{Pol-D-2}\eea

These polarization vectors possess the following properties
\bea &&\epsilon^{\pm}\cdot q=\epsilon^{\pm}\cdot
\ell=\epsilon^{\pm}\cdot
\epsilon^{\pm}=0~~,~~\epsilon^+\cdot\epsilon^-=-1~,~~~\nonumber\\
&&\epsilon^{L}\cdot
\epsilon^{T}=\spab{q|{\ell}|q}~~,~~\epsilon^{S_a}\cdot
\epsilon^{L,T}=0~~,~~\epsilon^{S_a}\epsilon^{S_b}=-\delta^{ab}~.~~~\label{inner-product}\eea

The metric can be decomposed as
\bea
\eta_{\mu\nu}={\epsilon^{L}\epsilon^{T}+\epsilon^{T}\epsilon^{L}\over\spab{q|{\ell}|q}}
-\epsilon^{+}\epsilon^{-}-\epsilon^{-}\epsilon^{+}-\sum_{a=1}^d\epsilon^{S_a}\epsilon^{S_a}~.~~~~~\label{metric-decom}\eea

The convention above 
can be generalized to $(4-2\epsilon)$-dimension trivially,
where we should replace $d\to -2\epsilon$.

To avoid notation confusion for the symbol ~$\widehat{}$~, an important convention is that we use  ~$\widehat{}$~ in two distinguished situations:  to represent extra-dimensional momenta $\widehat{\ell_i}$ and to represent BCFW shifted legs. Since we denote the leg in the scattering amplitude with momenta  $\widehat{\ell_i}$ by $\ell_i$ during the BCFW recursion and never do BCFW shift to it, this two situations are not hard to distinguish.

In our cases of amplitudes, we have two spin-1 extra-dimensional particles denoted by their 4-dimensional momenta $\ell_2$ and $\ell_1$.
Since there are $D-2$ helicity states $+~-$ and $S_a$ for each spin-1 particles with extra-dimensional momenta and all gluons have positive helicity, there are 9 kinds of amplitudes distinguished by the helicity of the two extra-dimensional particles. We denote these tree amplitudes as $\{\ell_2^+,\ell_1^{+}\}$, $\{\ell_2^+,\ell_1^{S_a}\}$ $\{\ell_2^+,\ell_1^{-}\}$, $\{\ell_2^-,\ell_1^{+}\}$, $\{\ell_2^-,\ell_1^{S_a}\}$, $\{\ell_2^{S_a},\ell_1^{+}\}$, $\{\ell_2^{S_a},\ell_1^-\}$, $\{\ell_2^-,\ell_1^-\}$, $\{\ell_2^{S_a}$, $\ell_1^{S_b}\}$.

\paragraph*{One 4-dimensional formulation }

 In this part, we review the 4-dimensional formulation of D-dimensional particles in \cite{Maitre:2007jq} , as an example to compare with the $D$-dimensional formulation above.

In \cite{Maitre:2007jq}, $D$-dimensional momentum $\bar \ell$ of mass $m$ is decomposed as below
\begin{align}
\bar \ell = \ell + \tilde \ell \, , \qquad \bar \ell^2  = \ell^2 -\mu^2 = m^2  \,  ,
\label{Eq:Dec0}
\end{align}
and its four-dimensional component $\ell$ are expressed as 
\begin{align}
\ell = \ell^\flat + \hat q_\ell \, , \qquad \hat q_\ell \equiv \frac{m^2+\mu^2 }{2\, \ell \cdot q_\ell} q_\ell  \,  ,
\label{Eq:Dec}
\end{align}
in terms of the two massless momenta $\ell^\flat$ and $q_\ell$, which is similar to (\ref{ell-bot}). For massive spin-$1\over{2}$ particles, they introduce tachyonic spinors for the degrees of freedom of extra-dimension, while for massless spin-1 particles, the degree of freedom is decomposed as 4-dimensional massive vector bosons' part in terms of $\{\varepsilon_{\pm}, \varepsilon_{0}\}$  and scalar part represented by factor $\hat G^{AB}$. The metric is decomposed as
\begin{eqnarray}
\sum_{i=1}^{D-2} \, \varepsilon_{i\, (d)}^\mu\left (\bar \ell , \bar \eta \right )\varepsilon_{i\, (d)}^{\ast \nu}\left (\bar \ell , \bar \eta \right ) &=&\left (   - \eta^{\mu \nu}  +\frac{ \ell^\mu \ell^\nu}{\mu^2} \right) \nonumber \\
&-&\left (  \tilde \eta^{\mu \nu}  +\frac{ \tilde \ell^\mu \tilde \ell^\nu}{\mu^2} \right ) \, ,
\label{Eq:CompGD2}
\end{eqnarray}
where
 \begin{align}
 -\eta^{\mu\nu}+\frac{\ell^{\mu}\ell^{\nu}}{\mu^{2}} &=  \sum_{\lambda=\pm,0}\varepsilon_{\lambda}^{\mu}(\ell) \, \varepsilon_{\lambda}^{*\nu}(\ell)  \, ,  \label{flat}
 \end{align}
in another kind of choices for 3 polarization vectors of a vector boson of mass $\mu$
\begin{align}
\varepsilon_{+}^{\mu}\left(\ell \right)    &= -\frac{\left[\ell^{\flat}\left|\gamma^{\mu}\right|  \hat q_\ell \right\rangle }{\sqrt{2}\mu} \, ,  &  &
\varepsilon_{-}^{\mu}\left(\ell \right)    = - \frac{\left\langle \ell^{\flat}\left|\gamma^{\mu}\right| \hat q_\ell \right]}{\sqrt{2}\mu}\, ,    \nn      
\varepsilon_{0}^{\mu}\left(\ell   \right)  &=  \phantom{-} \frac{\ell^{\flat\mu}-\hat q_\ell^{\mu}}{\mu} \, ,
\label{emass1}
\end{align}
and extra-dimensional part as
\begin{equation}
 \tilde \eta^{\mu \nu}  +\frac{ \tilde \ell^\mu \tilde \ell^\nu}{\mu^2}  \quad \to  \quad  \hat{G}^{AB} \equiv  G^{AB} - Q^A Q^B  \, .
 \label{Eq:Pref}
\end{equation}

This decomposition is typical to treat the $D-2$ degrees of freedom, which is useful in General Unitarity Method for 1-loop amplitudes to take the cut of propagator of the loop as massive vector bosons parts and scalar parts.

Different from (\ref{emass1}), $\epsilon^{\pm}$ in (\ref{Pol-D-1}) are equivalent to massive spin-1 states in \cite{Ferrario:2006np,Craig:2011ws,Kiermaier:2011cr} and we no longer have $\epsilon^{0}$ because we define the rest $D-4$ physical polarization vectors $\epsilon^{S_a}$ to be extra-dimensional, each equivalent to a massive scalar in \cite{Ferrario:2006np,Craig:2011ws,Kiermaier:2011cr} instead of (\ref{Eq:Pref}). So the tree amplitudes below are comparable with the massive results in \cite{Ferrario:2006np,Craig:2011ws,Kiermaier:2011cr}.

However, in our paper, the motivation to use those helicity choices $+~-$ and $S_a$ is that they are useful to express $D$-dimensional tree-level helicity amplitudes as inputs for $\mathcal{Q}$-cut representation and apply the BCFW recursion relations. Since we directly use the extra-dimensional polarization vectors and we can conveniently characterize the longitude degree of freedom by different helicity states.

\subsection{3-point amplitudes}
Since the 3-point amplitudes perform as building blocks in the BCFW recursion relations, we should calculate 3-point amplitudes first. Here we use simple Feynman rules to obtain
\bea
\cA(1,2,3)&=-\sqrt{2}\,\big[(p_2\cdot\epsilon_3)(\epsilon_1\cdot\epsilon_2)
+(p_3\cdot\epsilon_1)(\epsilon_2\cdot\epsilon_3)+(p_1\cdot\epsilon_2)(\epsilon_3\cdot\epsilon_1)\big]~.~~~
\label{3pointfey}
\eea

To use the BCFW recursion relations in our cases, we only need these 3-point amplitudes:  $\cA_3(1^{+},\ell_2^{+},\ell_1^{+})$, $\cA_3(1^{+},\ell_2^{+},\ell_1^{-})$, $\cA_3(1^{+},\ell_2^{+},\ell_1^{S_a})$, $\cA_3(1^{+},\ell_2^{-},\ell_1^{-})$, $\cA_3(1^{+},\ell_2^{-},\ell_1^{S_a})$, $\cA_3(1^+,\ell_2^{S_a},\ell_1^{S_b})$. Here we choose $1$ to represent the particle in 4-dimension and $\ell_1$, $\ell_2$ to represent the particles in D-dimension. Then we use the transverse polarization vectors given in (\ref{Pol-D-1}) to give specific expressions. When we use (\ref{3pointfey}) , we must pay attention to $(p\cdot\epsilon)$. If the momentum p has extra-dimensional components and $\epsilon=\epsilon^{S_a}$ ($a=1,2,\ldots, d$), the product  gets contribution from d-dimensional momentum component: $\epsilon^{S_a}(\widehat{\ell_1},q)\cdot p_{\widehat{\ell_2}}=\frac{\mu_{\ell_1,a} \Spab{q|\ell_2^{\bot}|q} }{\Spab{q|\ell_1^{\bot}|q}}-\mu_{\ell_2,a}$. 

The expressions for 3-point amplitudes are
\bea
&&\cA_3(1^{+},\ell_2^{+},\ell_1^{+})=\cA_3(1^{+},\ell_2^{+},\ell_1^{S_a})=0~,~~~\nonumber\\
&&\cA_3(1^{+},\ell_2^{+},\ell_1^{-})={\spbb{1~\ell_2^\bot}^3\over\spbb{\ell_2^\bot~\ell_1^\bot}\spbb{\ell_1^\bot~1}}~~~,~~~~~~~~~~~
\cA_3(1^{+},\ell_2^{-},\ell_1^{-})={\spaa{\ell_2^\bot~\ell_1^\bot}^3\over\spaa{\ell_1^\bot~1}\spaa{1~\ell_2^\bot}} ~,~~~\nonumber\\
&&{\cA_3(1^{+},\ell_2^{-},\ell_1^{S_a})={\sqrt{2}\mu_{\ell_1,a}{\spaa{\ell_2^\bot~q}\spaa{\ell_1^\bot~\ell_2^\bot}^2}\over
\spaa{\ell_1^\bot~q}\spaa{1~\ell_1^\bot}\spaa{1~\ell_2^\bot}}}~~,~~
{\cA_3(1^{+},\ell_2^{S_a},\ell_1^{-})=-{\sqrt{2}\mu_{\ell_2,a}{\spaa{\ell_1^\bot~q}\spaa{\ell_1^\bot~\ell_2^\bot}^2}\over
\spaa{\ell_2^\bot~q}\spaa{1~\ell_1^\bot}\spaa{1~\ell_2^\bot}}}~,~~~
\nonumber\\
&&\cA_3(1^{+},\ell_2^{S_a},\ell_1^{S_b})=-\delta^{ab}{\spbb{1~\ell_2^\bot}\spbb{1~\ell_1^\bot}\over\spbb{\ell_1^\bot~\ell_2^\bot}}~.~~~\label{3point}\eea

Notice that the reflection property of ${\cA_3(1^{+},\ell_2^{S_a},\ell_1^{-})}=(-1)^3\cA_3(1^{+},\ell_1^{-},\ell_2^{S_a})$, which is not obvious, is also obeyed because $\mu_{\ell_1,a}=-\mu_{\ell_2,a}$.

 \subsection{4-point amplitudes}
In appendix of \cite{Huang:2015cwh}, the expressions below are calculated using Feynman rules
\bea
&&\cA_{4}(1^+, 2^+,\ell_2^{+},\ell_1^{+})=\cA_{4}(1^+, 2^+,\ell_2^{+},\ell_1^{S_a})=0~,~~~
\nonumber\\
&&\cA_4(1^+,2^+,\ell_2^+,\ell_1^{-})={\mu^2
\spaa{\ell_1^\bot~q}^2\over\spaa{\ell_2^\bot~q}^2}{\spbb{2~1}\over\spaa{1~2}\spab{1|{\ell}_1|1}}~,~~~
\nonumber\\
&&\cA_4(1^+,2^+,\ell_2^{-},\ell_1^{-})=-{\spaa{\ell_1^\bot~\ell_2^\bot}^2}{\spbb{2~1}\over\spaa{1~2}\spab{1|{\ell}_1|1}}~,~~~
\nonumber\\
&&\cA_4(1^+,2^+,\ell_2^{-},\ell_1^{S_a})={\sqrt{2}\mu_{\ell_1,a}\spaa{\ell_1^\bot~\ell_2^\bot}\spaa{\ell_2^\bot~q}\over\spaa{\ell_1^\bot~q}}{\spbb{2~1}\over
\spaa{1~2}\spab{1|{\ell}_1|1}}~,~~~
\nonumber\\
&&\cA_4(1^+,2^+,\ell_2^{S_a},\ell_1^{-})=-{{\sqrt{2}\mu_{\ell_2,a}\spaa{\ell_2^\bot~\ell_1^\bot}\spaa{\ell_1^\bot~q}\over\spaa{\ell_2^\bot~q}}}{\spbb{2~1}\over
\spaa{1~2}\spab{1|{\ell}_1|1}}~,~~~
\nonumber\\
&&\cA_4(1^+,2^+,\ell_2^{S_a},\ell_1^{S_b})=-{\mu^2}{\spbb{2~1}\over\spaa{1~2}\spab{1|{\ell}_1|1}}\delta^{ab}~.~~~
\label{A4}
\eea

We have checked $\cA_4$ with the BCFW recursion relations. These expressions inspire us to guess the general expression for $\cA_n$ for $n\ge4$. The notations here are the same as in the following subsection \S\ref{general}.

\subsection{General expressions for the n-point amplitudes}
\label{general}
To express n-point tree amplitudes, we denote k positive helicity gluons as $1,2,\dots,k $, and 2 adjacent massless particle with extra-dimensional momenta by their 4-dimensional momenta ${\ell}_1,\ell_2$. (In this paper, we also denote this series of amplitudes by $\{\ell_2^{h_2},\ell_1^{h_1}\}$  for short)

Using the BCFW recursion relations, we find $n=k+2$ tree amplitudes do have general expressions  for $n\ge4$. Here we list our main results
\bea
\cA_{n}(1^+, 2^+,...,k^+, \ell_2^{+},\ell_1^{+})&=&\cA_{n}(1^+, 2^+,...,k^+, \ell_2^{+},\ell_1^{S_a})=0~,~~~
\nonumber\\
\cA_{n}(1^+, 2^+,...,k^+, \ell_2^{+},\ell_1^{-})&=&
 -{{\mu^2 
\spaa{q~\ell_1^{\bot}}^2}
\over
{\spaa{q~\ell_2^{\bot}}^2}}
\times
{\spbb{1|\prod_{i=2}^{k-1}(\mu^2-x_{i,\ell_1}x_{\ell_1,i+1})|k}
\over 
\spaa{1~2}\cdots \spaa{k-1~k}\prod_{i=2}^{k}(\mu^2+x_{\ell_1,i}^2)}~,~~~
\nonumber\\
\cA_{n}(1^+, 2^+,...,k^+, \ell_2^{-},\ell_1^{-})&=&
{\spaa{\ell_1^\bot~\ell_2^\bot}^2}
\times
{\spbb{1|\prod_{i=2}^{k-1}(\mu^2-x_{i,\ell_1}x_{\ell_1,i+1})|k}
\over \spaa{1~2}\cdots \spaa{k-1~k}\prod_{i=2}^{k}(\mu^2+x_{\ell_1,i}^2)}~,~~~
\nonumber\\
\cA_{n}(1^+, 2^+,...,k^+, \ell_2^{-},\ell_1^{S_a})&=&
-{\sqrt{2}\mu_{\ell_1,a}
\spaa{\ell_1^\bot~\ell_2^\bot}\spaa{\ell_2^\bot~q}\over
\spaa{\ell_1^\bot~q}}
\times
{\spbb{1|\prod_{i=2}^{k-1}(\mu^2-x_{i,\ell_1}x_{\ell_1,i+1})|k}
\over \spaa{1~2}\cdots \spaa{k-1~k}\prod_{i=2}^{k}(\mu^2+x_{\ell_1,i}^2)}~,~~~
\nonumber\\
\cA_{n}(1^+, 2^+,...,k^+, \ell_2^{S_a},\ell_1^{-})&=&
{\sqrt{2}\mu_{\ell_2,a}
\spaa{\ell_2^\bot~\ell_1^\bot}\spaa{\ell_1^\bot~q}\over
\spaa{\ell_2^\bot~q}}
\times
{\spbb{1|\prod_{i=2}^{k-1}(\mu^2-x_{i,\ell_1}x_{\ell_1,i+1})|k}
\over \spaa{1~2}\cdots \spaa{k-1~k}\prod_{i=2}^{k}(\mu^2+x_{\ell_1,i}^2)}~,~~~
\nonumber\\
\cA_{n}(1^+, 2^+,...,k^+, \ell_2^{S_a},\ell_1^{S_b})&=&
\mu^2 \delta^{ab}
\times
{\spbb{1|\prod_{i=2}^{k-1}(\mu^2-x_{i,\ell_1}x_{\ell_1,i+1})|k}
\over \spaa{1~2}\cdots \spaa{k-1~k}\prod_{i=2}^{k}(\mu^2+x_{\ell_1,i}^2)}~,~~~
\label{result}
\eea
where q is a null 4-dimensional reference momentum and $q\cdot \ell_i\neq 0$, $\ell_i^{\bot}=\ell_i-{\vec{\mu}^2\over
\Spab{q|\ell_i|q}}q$, $\mu^2$ is the inner product of all the extra dimensional momentum values of two $\widehat{\ell_i}$, while $\mu_{\ell_i,a}$ is the extra-dimensional momentum component of $\widehat{\ell_i}$ in the direction $\vec{e}_a$ of the extra-dimensional polarization vector $\epsilon^{S_a}(\ell_{i})$. We borrow the notations $x_{i, \ell_1}=p_i+p_{i+1}+\dots +p_k+p_{\ell_2}$ and  $x_{\ell_1,i}=p_{\ell_1}+p_{1}+\dots +p_{i-1}$ from \cite{Craig:2011ws} and set${\spbb{1|\prod_{i=2}^{k-1}(\mu^2-x_{i,\ell_1}x_{\ell_1,i+1})|2}}=\spbb{1~2}$ for n=4 (k=2), so these expressions also hold for $\cA_4$. Massive versions of $\{\ell_2^{S_a},\ell_1^{S_b}\}$  and $\{\ell_2^+,\ell_1^{-}\}$ have been obtained in \cite{Ferrario:2006np} and \cite{Craig:2011ws} separately but the other two $\{\ell_2^-,\ell_1^{-}\}$ and $\{\ell_2^-,\ell_1^{S_a}\}$ are new here.

One observation is that these amplitudes have a common structure: they all have one common part ${\spbb{1|\prod_{i=2}^{k-1}(\mu^2-x_{i,\ell_1}x_{\ell_1,i+1})|k}
\over \spaa{1~2}\cdots \spaa{k-1~k}\prod_{i=2}^{k}(\mu^2+x_{\ell_1,i}^2)}$ changing with the increasing number of k, while the other part, stays the same for any number of k and depends on $\ell_1^{h_1},\ell_2^{h_2}, q, \mu_{l_i,a}, \mu^2$. This structure is astonishingly simple!


\subsection{Structure of the general exprssions}
The general expressions for n-points amplitudes with $n>4$ are not easy to get from Feynman-diagrams. Here based on our assumption that they have a common structure inspired by 4-point cases (\ref{A4}), we build a clever proof by using the BCFW recursion relations.

Let's study the common structure first. We can see $\cA_n$ in (\ref{result}) have two parts, we denote them as $PartA(\ell_1^{h_1},\ell_2^{h_2})$ and $PartB(k)$.

$PartB(k)$ is same for all non-zero cases
\bea
PartB(k)= {\spbb{1|\prod_{i=2}^{k-1}(\mu^2-x_{i,\ell_1}x_{\ell_1,i+1})|k}
\over \spaa{1~2}\cdots \spaa{k-1~k}\prod_{i=2}^{k}(\mu^2+x_{\ell_1,i}^2)}~,~~~
\eea
while $PartA$ is different
\bea
PartA(\ell_1^{+},\ell_2^{-})&=& -{{\mu^2 
\spaa{q~\ell_1^{\bot}}^2}
\over
{\spaa{q~\ell_2^{\bot}}^2}}~,~~~
\nonumber\\
PartA(\ell_1^{-},\ell_2^{-})&=&{\spaa{\ell_1^\bot~\ell_2^\bot}^2}~,~~~
\nonumber\\
PartA(\ell_1^{-},\ell_2^{S_a})&=&-{\sqrt{2}\mu_{\ell_1,a}
\spaa{\ell_1^\bot~\ell_2^\bot}\spaa{\ell_2^\bot~q}\over
\spaa{\ell_1^\bot~q}}~,~~~
\nonumber\\
PartA(\ell_1^{S_a},\ell_2^{-})&=&{\sqrt{2}\mu_{\ell_2,a}
\spaa{\ell_2^\bot~\ell_1^\bot}\spaa{\ell_1^\bot~q}\over
\spaa{\ell_2^\bot~q}}~,~~~
\nonumber\\
PartA(\ell_1^{S_a},\ell_2^{S_b})&=& \mu^2 \delta_{ab}~,~~~
\eea
and we can see $PartA(\ell_2^{h},\ell_1^{h_1})$ is independent of all k gluons.

The pole structure of these $\cA_n$ is contained in $PartB(k)$. We see there are not only 2-particle poles $\spaa{1~2}\cdots \spaa{k-1~k}$ in $PartB(k)$ but also multi-particle poles $\prod_{i=2}^{k}(\mu^2+x_{\ell_1,i}^2)$.

\section{Proof by Mathematical Induction}
\label{proof}

In this section, we give our proof  of the results (\ref{result}) by mathematical induction. First we choose a special shift so that the BCFW recursion relations for these tree amplitudes can be reduced much simple. Then we separate our proof for the inductive step into two parts. 

As the first step of a mathematical induction, known as the base case, we can see $\cA_4$ amplitudes given in (\ref{A4}) match the expressions (\ref{result}) for $n=4$.

The second step is called the inductive step. Let's assume the expressions (\ref{result}) hold for n-point cases. If the expressions (\ref{result}) hold for (n+1)-point cases, we can conclude that (\ref{result}) are general expressions for $n\ge4$. This hypothesis the expressions (\ref{result}) hold for (n+1)-point cases, is called the inductive hypothesis. By proving the hypothesis, we can prove the expressions hold for any natural number n for $n\ge4$.

\subsection{Strategies to use the BCFW recursion relations}
Using BCFW recursion relations \cite{Britto:2004ap,Britto:2005fq}, one can get higher-point tree amplitudes from lower ones.
An instruction to use the BCFW recursion relations and common notations can be found in textbooks such as \cite{Elvang:2013cua}.

After shifting the momentum with complex parameter z, tree amplitudes without boundary contributions can be calculated by
\begin{equation}
\label{BCFW}
A_{n}=\sum_{\text{i, j}} \sum_{h}  \frac{A_i^{h}(z_{ij})A_{j}^{-h}(z_{ij})}{P(0)^2}~.~~~
\end{equation}
where we need to sum all possible partitions of subdiagrams noted by i, j as well as all possible helicity $h$ channels with solutions $z=z_{ij}$ putting the propagator $P(z)^2=0$ to get $A_i$ and $A_j$ on-shell. In the following, we will show how to use the recursion relations explicitly in our proof.


In our cases, the intermedia particle have extra-dimensional momentum, so we need to sum over the intermedia particle's helicity $+ - S_a$ rather than $+ -$ in 4-D cases.

To calculate $\cA_{n+1}$  (n=k+2) from lower-point amplitudes, we choose the BCFW shift to be $[k|k+1\rangle^+$
\bea
  |\WH{k}] = |k] - z\,|k+1]\,,   
  ~~~~~~
  |\WH{k}\rangle = |k\rangle\,,   
  ~~~~~~
  |\WH{k+1}] = |k+1]\,,
  ~~~~~~
  |\WH{k+1}\rangle = |k+1\rangle + z |k\rangle \,.
\eea

This shift involving two adjacent gluons mainly have two advantages.
One is that poles $\spaa{1~2}\cdots\spaa{k-1~k}$ and $\prod_{i=2}^{k}(\mu^2+x_{\ell_1,i}^2)$ in the denominator of $PartB(k)$ would not change after the shift.

\begin{figure}[h]
\centering
\includegraphics[scale=0.6]{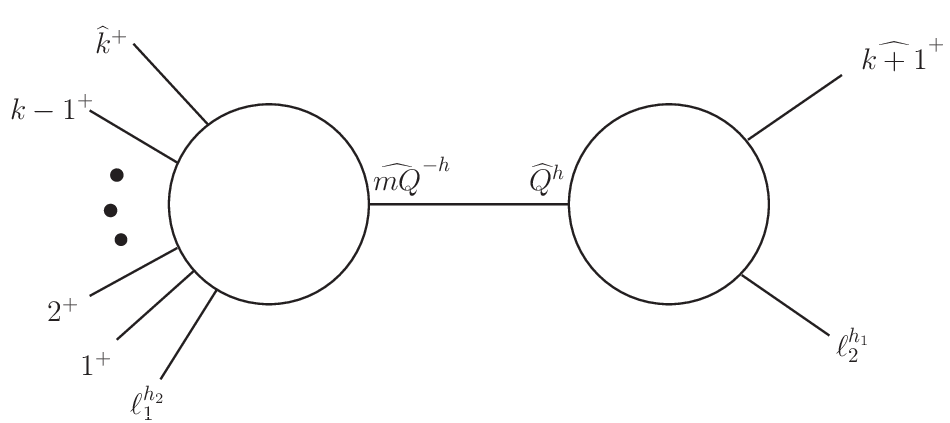}
\caption{The general situation for $\{\ell_2^{h_1},\ell_1^{h_2}\}$}
\end{figure}

The other advantage is that we can reduce (\ref{BCFW}) to
\bea
\cA_{n+1}(1^+,2^+,\cdots ,(k+1)^+,\ell_2^{h_2},\ell_1^{h_1})&=&\sum_{h}\frac{A_L(1^+,2^+,...,\widehat{k}^+,\widehat{mQ}^{-h},\ell_1^{h_1})\times A_R(\widehat{k+1}^+,\ell_2^{h_2},\widehat{Q}^{h})}{P(0)^2}~,~~~
\label{reduced-BCFW}
\eea 
where $P_{\widehat{mQ}}=-P_{\widehat{Q}}=P_{\widehat{k+1}}+P_{\widehat{\ell_2}}$ for momentum conservation. Symbol ~$\widehat{}$~ above the legs $1\cdots k+1$, Q and mQ in (\ref{reduced-BCFW}) stands for the leg being on-shell after the shift and we denote legs which have extra-dimension momenta $\WH{\ell_i}$ by $\ell_i$,  since we never do BCFW shift to $\ell_i$. We see one only needs to replace $\ell_2$ in $A_{n}$ with $\widehat{mQ}$ to express $A_L$.

The reason why we can obtain (\ref{reduced-BCFW}) is as follows. When both of $\ell_1$ and $\ell_2$ are in the same diagram $A_L$ or $A_R$, the intermedia massless particles denoted as $\widehat{Q}$ and $\widehat{mQ}$ will have no extra-dimensional momentum component. One can find the other diagram which has no leg with extra-dimensional momentum always vanishes. In a conclusion, a non-vanishing $A_L\times A_R$ term should be a product of a n-point amplitude and a 3-point amplitude as (\ref{reduced-BCFW}).

For zero cases in (\ref{result}), using (\ref{reduced-BCFW}) we find are they do vanish because Right Hand Side of (\ref{reduced-BCFW}) of $\{\ell_2^+,\ell_1^+\}, \{\ell_2^+,\ell_1^{S_a}\}, \{\ell_2^{S_a},\ell_1^+\}$ have at least one vanishing lower-point amplitude of $A_L$  and $A_R$.

Finally, let's focus on non-zero cases. Our strategy for the proof is to deal with $PartA$ and $PartB$ separately  in our recursion from $n$ to $n+1$. We rewrite (\ref{reduced-BCFW}) as

\bea
\cA_{n+1}(1^+,2^+,\cdots,(k+1)^+, \ell_2^{h_1},\ell_1^{h_2})&=&\sum_{h}\frac{A_L(1^+,2^+,...,\widehat{k}^+,\widehat{mQ}^{-h},\ell_1^{h_2})\times A_R(\widehat{k+1}^+,\ell_2^{h_1},\widehat{Q}^{h})}{P(0)^2}
\nonumber\\
&=&{PartB(\widehat{k})\sum_{h}PartA(\widehat{mQ}^{h},\ell_1^{h_1}) \times A_3(\widehat{k+1}^+, \widehat{Q}^{-h},\ell_2^{h_2})\over{P(0)^2}}
~.~~~
\eea

With these notations, we separate the proof to two parts. The first part is
\bea
\sum_{h}PartA(\widehat{mQ}^{h},\ell_1^{h_1}) \times A_3(\widehat{k+1}^+, \widehat{Q}^{-h},\ell_2^{h_2})
&\rightarrow& PartA(\ell_2^{h},\ell_1^{h_1}) {\Spab{k|\ell_2|k+1} \over{\Spaa{k~k+1}}}~,~~~
\label{PartA}
\eea
while the second part is:
\bea
PartB(\widehat{k})= {\spbb{1|\prod_{i=2}^{k-1}(\mu^2-x_{i,\ell_1}x_{\ell_1,i+1})|\widehat{k}}
\over \spaa{1~2}\cdots \spaa{k-1~k}\prod_{i=2}^{k}(\mu^2+x_{\ell_1,i}^2)}
&\rightarrow&
  {\spbb{1|\prod_{i=2}^{k}(\mu^2-x_{i,\ell_1}x_{\ell_1,i+1})|k+1}
\over \spaa{1~2}\cdots \spaa{k-1~k}\prod_{i=2}^{k}(\mu^2+x_{\ell_1,i}^2)} 
{1\over\Spab{k|\ell_2|k+1}}
\nonumber\\
&=&PartB(k+1){\spaa{k~k+1}P(0)^2\over\Spab{k|\ell_2|k+1}}
~.~~~
\label{PartB}
\eea

According to the partition in (\ref{reduced-BCFW}), where two legs (k+1) and $\ell_2$ are always on the right side, we can rewrite $P(0)^2$
\begin{equation}
P(0)^2=(p_{\widehat{\ell_2}}+ p_{k+1})^2=(p_{\ell_1}+p_1+p_2+..+p_{k})^2+\mu^2~.~~~
\end{equation}

Define $x_{\ell_1,i}=p_{\ell_1}+p_1+p_2+..+p_{i-1}$, then

\begin{equation}
P(0)^2=\mu^2+x_{\ell_1,k}^2~.~~~
\end{equation}

The on-shell condition for $P(z)^2=0$ is $P_{\widehat{Q}}^2=(P_{\widehat{k+1}}+P_{\widehat{\ell_2}})^2=0$, with the solution for z
\bea
z=z_L(k+1) &= &-{{\langle k+1|\ell_2|k+1]}\over{\langle \text{k}|\ell_2|k+1]}}~.~~~
\eea


\subsection{Proof of the inductive hypothesis}
Here we prove the inductive hypothesis by showing that (\ref{PartA}) and (\ref{PartB}) are correct.

The proof for (\ref{PartB}) is easy.

After the shift, $PartB(k)$ changes with $|\widehat{k}]$
\bea |\widehat{k}]=|k]-z_L(k+1)|k+1]=|k]+{{\langle k+1|\ell_2|k+1]}\over{\langle k|\ell_2|k+1]}}~|k+1]={\mu^2-x_{k,\ell_1}x_{\ell_1,k+1}
\over{\langle k|\ell_2|k+1]}}|k+1]~,~~~
\label{shiftk}
\eea
where we use identity $|k|\ell_2|k+1]+|k+1|\ell_2|k+1]=|\mu^2|k+1]-|(p_k+p_{k+1}+p_{\ell_2})|\ell_1+1+\cdots+k|k+1]$

Applying (\ref{shiftk}) to $PartB(k)$, we see (\ref{PartB}) is true, where we can see one advantage of the special BCFW shift we choose. 

However, the proof for  (\ref{PartA}) is very difficult. 

The inductive steps (\ref{PartA}) are different for each non-zero cases. Regarding the results for $\{\ell_2^+,\ell_1^+\}$, $\{\ell_2^+,\ell_1^{S_a}\}$, $\{\ell_2^{S_a},\ell_1^+\}$ cases are zero, the remaining non-zero cases are $\{\ell_2^+,\ell_1^{-}\}$, $\{\ell_2^-,\ell_1^{+}\}$, $\{\ell_2^-,\ell_1^{S_a}\}$, $\{\ell_2^{S_a},\ell_1^-\}$,  $\{\ell_2^-,\ell_1^-\}$, $\{\ell_2^{S_a},\ell_1^{S_b}\}$. We only need to calculate 4 of them: $\{\ell_2^+,\ell_1^{-}\},\{\ell_2^-,\ell_1^{S_a}\},~\{\ell_2^{S_a},\ell_1^{S_b}\},~\{\ell_2^{-},\ell_1^{-}\}$. 

Recall (\ref{PartA}) is to show
\bea
\sum_{h}PartA(\widehat{mQ}^{h},\ell_1^{h_1}) \times A_3(\widehat{k+1}^+, \widehat{Q}^{-h},\ell_2^{h_2})={PartA(\ell_2^{h_2}, \ell_1^{h_1})\times \Spab{k|\ell_2|k+1}\over \Spaa{k~k+1}}~.~~~
\label{PartA3p}
\eea

The proof for (\ref{PartA3p}) is nontrivial. We conclude the proof as the following steps:
\begin{itemize}
\item
First, substitute all the spinor brackets and angles of $Q^{\bot}$ with associated spinor brackets and angles of $mQ^{\bot}$ due to the momentum conservation. Second, multiply  every $|\widehat{mQ}^{\bot}]$ (or $|\widehat{mQ}^{\bot}\rangle$) with $\spaa{\widehat{mQ}^{\bot}~ q}$ (or $\spbb{\widehat{mQ}^{\bot}~ q}$) so that we can can replace $\widehat{mQ}^{\bot}$ by $\widehat{mQ}$ for $\widehat{mQ}^{\bot}=\widehat{mQ}-{\vec{\mu}^2\over\Spab{q|\widehat{mQ}|q}}q$.

\item
Next, before simplifying the spinor products, if there is $\langle q|\widehat{mQ}|q]$ appearing in the denominator, we should eliminate all the  $\langle q|\widehat{mQ}|q]$ in the first place, by technicallly using the Schouten Identity to $\langle \ell_i^{\bot}|\widehat{mQ}|q]$ with another term including $~|q\rangle$.  We will specifically show this step for each case latter.
\item
Finaly, if there is no $\langle q|\widehat{mQ}|q]$ left, we begin to simplify the spinor products using the following identities
\bea
\spaa{q~\widehat{k+1}}&=&-{\spaa{k~k+1}\spbb{k+1~\ell_2^\bot}\spaa{\ell_2^\bot~q}
\over \spab{k|\ell_2|k+1}} ~,~~~
\nonumber\\
\spaa{\widehat{k+1}~\ell_2^\bot}&=&{\mu^2\spaa{k~k+1}\spbb{k+1~q}\over\spbb{\ell_2^\bot~q}\spab{k|\ell_2|k+1}}~,~~~
\nonumber\\
\spab{q|\widehat{mQ}|\ell_2^{\bot}}
&=&
\frac{\langle \ell_2^{\bot}~q\rangle  \langle k~k+1\rangle  [k+1~\ell_2^{\bot}]^2}{\langle k|\ell_2|k+1]}~,~~~
\nonumber\\
\langle \ell_2^{\bot}|\widehat{mQ}|q]
&=&
-\frac{\mu^2 \langle k~k+1\rangle  [q~k+1]^2}{\langle k|\ell_2|k+1][q~\ell_2^{\bot}]}~,~~~
\nonumber\\
\langle q|\widehat{mQ}|k+1]
&=&
\spab{q|\ell_2^{\bot}|k+1}~,~~~
\label{identities}
\eea

which is no hard to prove by substituting $\widehat{k+1}$, $\widehat{k}$ and $\widehat{mQ}$ then using the Schouten identity to combine the separate terms.

\end{itemize}

In the following, we will prove (\ref{PartA3p}) is true for each non-zero cases in detail. We have also verified (\ref{PartA3p}) numerically using S@M package\cite{Maitre:2007jq} for each case. 

\subsubsection{$\{\ell_2^+,\ell_1^{-}\}$ and $\{\ell_2^{S_a},\ell_1^{S_b}\}$}
Both of the two cases $\{\ell_2^+,\ell_1^{-}\}$ and $\{\ell_2^{S_a},\ell_1^{S_b}\}$ only have one contributing term and no $\langle q|\widehat{mQ}|q]$ will appear in the denominators. We only need to simplify all the spinor products.

For $\{\ell_2^+,\ell_1^{-}\}$

\begin{figure}[h]
\centering
\includegraphics[scale=0.8]{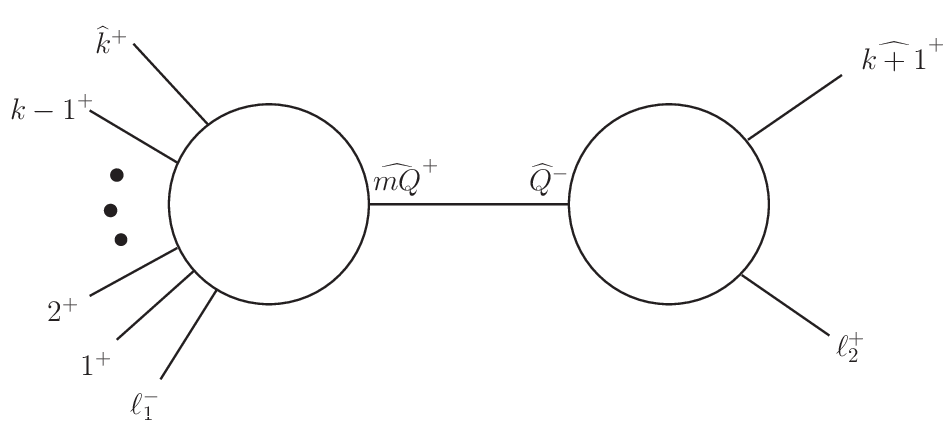}
\caption{$\{ \ell_2^+,\ell_1^- \}$}
\end{figure}

\bea 
&&PartA(\widehat{mQ}^{+},\ell_1^{-}) \times A_3(\widehat{k+1}, \widehat{Q}^{-},\ell_2^{+})
\nonumber\\
&=&
-{u^2 
\spaa{q~\ell_1^{\bot}}^2
\over \spaa{q~\widehat{Q}^{\bot}}^2}
\times
 \frac{[k+1 ~\ell_2^{\bot}]^3}{[k+1~\widehat{mQ}^{\bot}] [\widehat{mQ}^{\bot}~\ell_2^{\bot}]}
\nonumber\\ 
&=&
 -{u^2 
\spaa{q~\ell_1^{\bot}}^2
  [k+1 ~\ell_2^{\bot}]^3
 \over
 \langle q|\widehat{mQ}|k+1] \langle q|\widehat{mQ}|\ell_2^{\bot}]
 }
 \nonumber\\
 &=&
 {-{{\mu^2 \spaa{q~\ell_1^{\bot}}^2}\over{\spaa{q~\ell_2^{\bot}}^2}}}
\times
{ \Spab{k|\ell_2|k+1}\over \Spaa{k~k+1}}~,~~~
\eea
where for the last equality, we use the identities  in (\ref{identities}).

The case $\{\ell_2^{S_a},\ell_1^{S_b}\}$ is similar to $\{\ell_2^+,\ell_1^{-}\}$, we can easily get

\begin{figure}[h]
\centering
\includegraphics[scale=0.8]{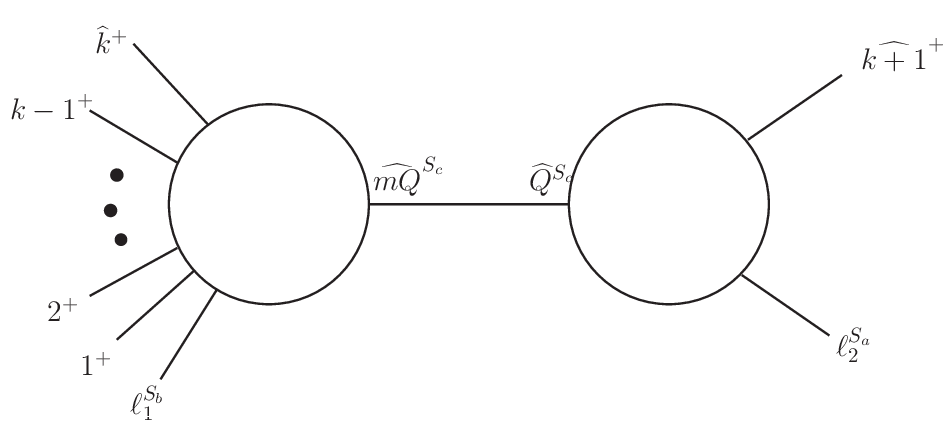}
\caption{$ \{\ell_2^{S_a},\ell_1^{S_b}\}$}
\end{figure}

\bea
&&\sum_{S_c}PartA(\widehat{mQ}^{S_c},\ell_1^{S_b}) \times A_3(\widehat{k+1}^+, \widehat{Q}^{S_c},\ell_2^{S_a})
\nonumber\\
&=&
\sum_{S_c}-{\mu^2}\delta_{cb}\times{(-\delta_{ac}{\spbb{k+1~\ell_2^\bot}\spbb{k+1~\WH{Q}^\bot}\over\spbb{\WH{Q}^\bot~\ell_2^\bot}})}
\nonumber\\
&=&
-{\mu^2}\delta_{ab}{\spbb{k+1~\ell_2^\bot}\spba{k+1|\WH{mQ}^\bot|q}\over\spab{q|\WH{mQ}^\bot|\ell_2^\bot}}
\nonumber\\
&=&
-{\mu^2}\delta_{ab}\times{ \Spab{k|\ell_2|k+1}\over \Spaa{k~k+1}}
\eea

\subsubsection{$\{\ell_2^{-},\ell_1^{S_a}\}$}

This case has two contributing terms

\begin{figure}[h]
\centering
{\includegraphics[scale=0.5]{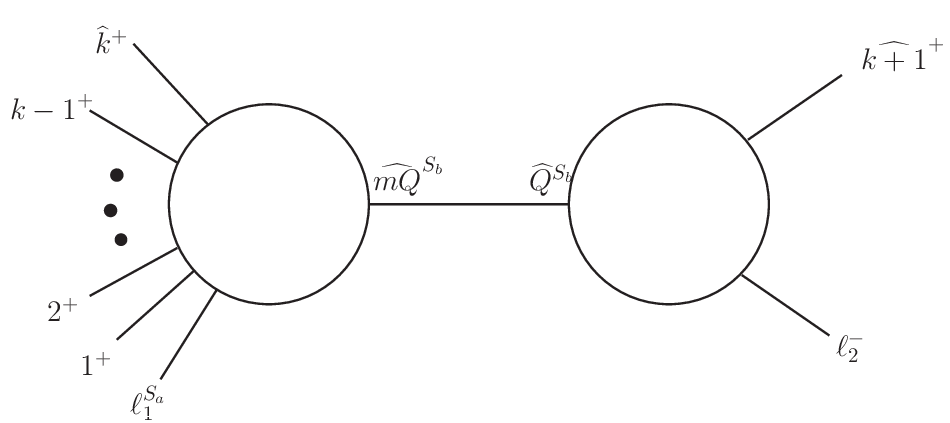}}
{\includegraphics[scale=0.5]{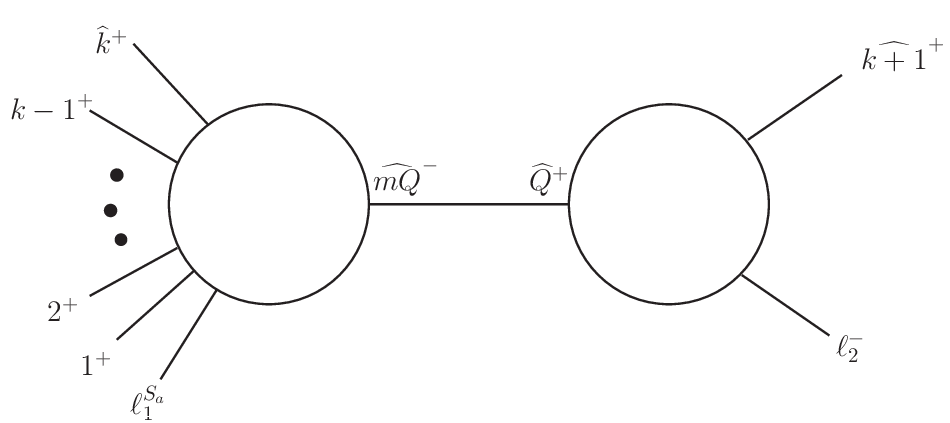}}
\caption{$\{\ell_2^-,\ell_1^{S_a}\}$}
\end{figure}

\bea
\sum_{h} PartA(\widehat{mQ}^h,\ell_1^{h_1})\times A_R(\widehat{k+1}^+,\ell_2^{h_2},\WH{Q}^{-h})
&=& PartA(\widehat{mQ}^-,\ell_1^{S_a})\times\cA_3(\widehat{k+1}^+,\ell_2^-,\widehat{Q}^+)~~~~
\nonumber\\
&+&\sum_{S_b} PartA(\widehat{mQ}^{S_b},\ell_1^{S_a})\times\cA_3(\widehat{k+1}^+,\ell_2^-,\widehat{Q}^{S_b})~.~~~
\eea

First we need to separate $\cA_3(1^{+},\ell_2^{-},\ell_1^{S_a})$ to two terms
\bea
{\cA_3(1^{+},\ell_2^{-},\ell_1^{S_a})={\sqrt{2}\mu_{\ell_1,a}{\spaa{\ell_2^\bot~q}\spbb{1~q}\over
\spaa{q~1}\spbb{q~\ell_2^\bot}}}}+{\sqrt{2}\mu_{\ell_1,a}{\spaa{\ell_2^\bot~q}^2\spbb{1~q}\over
\spaa{q~1}\spaa{\ell_1^{\bot}~q}\spbb{q~\ell_1^{\bot}}}}~.~~~
\eea

Then we write them explicitly
\bea
PartA(\widehat{mQ}^-,\ell_1^{S_a})\times\cA_3(\widehat{k+1}^+,\ell_2^-,\widehat{Q}^+)
&=&-\sqrt{2}\mu_{\ell_1,a}\spaa{\ell_1^\bot~\widehat{mQ}^\bot}\spaa{\widehat{mQ}^\bot~q}\spbb{\widehat{mQ}^\bot~k+1}^3
\over \spaa{\ell_1^\bot~q}\spbb{k+1~\ell_2^\bot}\spbb{\ell_2^\bot~\widehat{mQ}^\bot}~~~~
\nonumber\\
&=&-{\sqrt{2}\mu_{\ell_1,a}\spab{\ell_1^\bot|\widehat{mQ}^\bot|q}{\spab{q|\widehat{mQ}^\bot|k+1}^3}
\over \spaa{\ell_1^\bot~q}\spbb{k+1~\ell_2^\bot}\spab{q|\widehat{mQ}^\bot|\ell_2^\bot}\spab{q|\widehat{mQ}^\bot|q}}~,~~~
\eea
and
\bea
&&\sum_{S_b} PartA(\widehat{mQ}^{S_b},\ell_1^{S_a})\times\cA_3(\widehat{k+1}^+,\ell_2^-,\widehat{Q}^{S_b})
\nonumber\\
&=&\sum_{S_b}({\mu^2\delta_{ab}{\sqrt{2}\mu_{\ell_1,b}{\spaa{\ell_2^\bot~q}\spbb{k+1~q}\over
\spaa{q~\widehat{k+1}}\spbb{q~\ell_2^\bot}}}}+{\mu^2\delta_{ab}{\sqrt{2}\mu_{\ell_1,b}{\spaa{\ell_2^\bot~q}^2\spbb{k+1~q}\over
\spaa{q~\widehat{k+1}}\spaa{\widehat{mQ}^\bot~q}\spbb{q~\widehat{mQ}^\bot}}}})~~~~
\nonumber\\
&=&{\mu^2{\sqrt{2}\mu_{\ell_1,a}{\spaa{\ell_2^\bot~q}\spbb{k+1~q}\over
\spaa{q~\widehat{k+1}}\spbb{q~\ell_2^\bot}}}}-{\mu^2{\sqrt{2}\mu_{\ell_1,a}{\spaa{\ell_2^\bot~q}^2\spbb{k+1~q}\over
\spaa{q~\widehat{k+1}}\spab{q|\widehat{mQ}|q}}}}~,~~~
\eea

Instead of simplifing all the spinor products directly, the first thing to do is to apply the Schouten Identity to ${\langle \ell_1^{\bot}|\widehat{mQ}|q] \langle q|\widehat{mQ}|k+1]}$ to eliminate the unphysical pole $\langle q|\widehat{mQ}|q]$
\bea
{\langle \ell_1^{\bot}|\widehat{mQ}|q] \langle q|\widehat{mQ}|k+1]}={\langle \ell_1^{\bot}|\WH{mQ}|k+1] \langle q|\widehat{mQ}|q]}+\mu^2\spaa{\ell_1^{\bot}~q}\spbb{k+1~q}~,~~~
\label{l1k+1}
\eea
where $\spab{\ell_1^\bot|\widehat{mQ}|k+1}$ can be written as
\bea
\spab{\ell_1^\bot|\widehat{mQ}|k+1}
&=&\spab{\ell_1^\bot|\ell_2|k+1}~~~~
\nonumber\\
&=&\spab{\ell_1^\bot|\ell_2^\bot+{\mu^2\over\spab{q|\ell_2|q}}q|k+1}~~~~
\nonumber\\
&=&\spaa{\ell_1^\bot~\ell_2^\bot}\spbb{\ell_2^\bot~k+1}+{\mu^2\spaa{\ell_1^\bot~q}\spbb{q~k+1}
\over\spaa{q~\ell_2^\bot}\spbb{\ell_2^\bot~q}}~.~~~
\eea

Surprisingly the result can be merged and simplified

\bea
&&\sum_{h}PartA(\widehat{mQ}^{h},\ell_1^{S_a}) \times A_3(\widehat{k+1}, \widehat{Q}^{-h},\ell_2^{-})
\nonumber\\
&=&PartA(\widehat{mQ}^{-},\ell_1^{S_a}) \times A_3(\widehat{k+1}, \widehat{Q}^{+},\ell_2^{-})+PartA(\widehat{mQ}^{S_a},\ell_1^{S_a}) \times A_3(\widehat{k+1}, \widehat{Q}^{S_a},\ell_2^{-})
\nonumber\\
&=&{\sqrt{2}\mu_{\ell_1,a}\spab{k|\ell_2|k+1}\spab{\ell_1^\bot|\widehat{mQ}^\bot|q}
\spab{q|\widehat{mQ}^\bot|k+1}\over\spaa{\ell_1^\bot~q}\spbb{k+1~\ell_2^\bot}\spaa{k~k+1}
\spab{q|\widehat{mQ}^\bot|q}}-{\mu^2\sqrt{2}\mu_{\ell_1,a}\spbb{k+1~q}\spab{k|\ell_2|k+1}\over
\spaa{k~k+1}\spbb{k+1~\ell_2^\bot}\spbb{\ell_2^\bot~q}}
\nonumber\\
&+&{\mu^2\sqrt{2}\mu_{\ell_1,a}\spaa{\ell_2^\bot~q}\spbb{k+1~q}\spab{k|\ell_2|k+1}
\over\spaa{k~k+1}\spbb{k+1~\ell_2^\bot}}
\nonumber\\
&=&
{\sqrt{2}\mu_{\ell_1,a}
\spaa{\ell_2^\bot~\ell_1^\bot}\spaa{\ell_1^\bot~q}\spab{k|\ell_2|k+1}\over
\spaa{\ell_2^\bot~q}\spaa{k~k+1}}~.~~~
\eea

\subsubsection{$\{\ell_2^-,\ell_1^-\}$}
$\{\ell_2^-,\ell_1^-\}$ case has three contributing terms

\begin{figure}[h]
\centering
\subfloat[]{\includegraphics[scale=0.35]{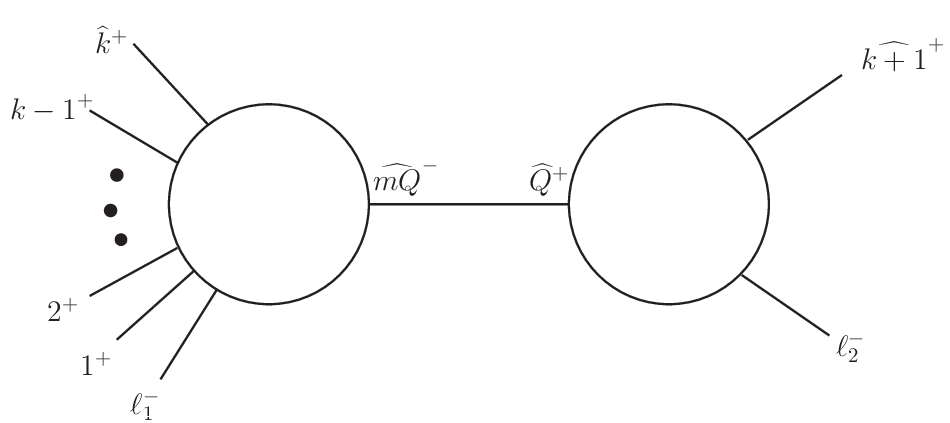}}
\subfloat[]{\includegraphics[scale=0.35]{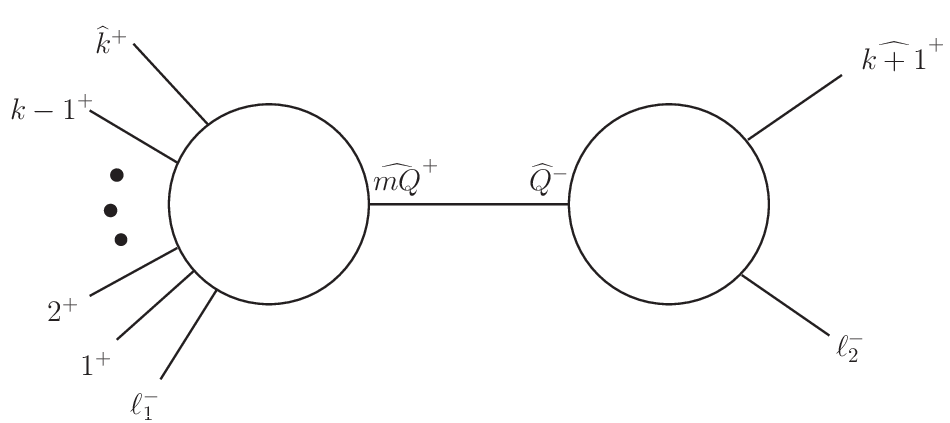}}
\subfloat[]{\includegraphics[scale=0.35]{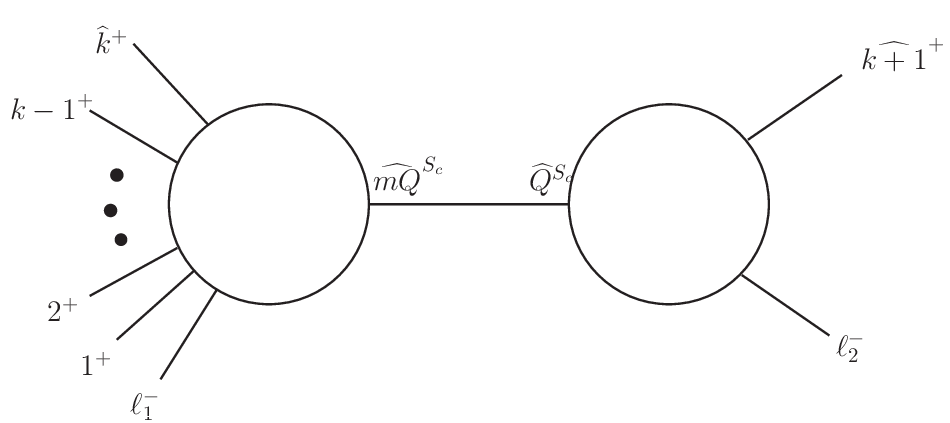}}
\caption{$\{\ell_2^-,\ell_1^-\}$}
\end{figure}

\bea
\sum_{h} PartA(\widehat{mQ}^h,\ell_1^{h_1})\times A_R(\widehat{k+1}^+,\ell_2^{h_2},m\WH{Q}^{-h})
&=& PartA(\widehat{mQ}^-,\ell_1^-)\times\cA_3(\widehat{k+1}^+,\ell_2^-,\widehat{Q}^+)~~~~
\nonumber\\
&+& PartA(\widehat{mQ}^+,\ell_1^-)\times\cA_3(\widehat{k+1}^+,\ell_2^-,\widehat{Q}^-)~~~~
\nonumber\\
&+&\sum_{Sc} PartA(\widehat{mQ}^{Sc},\ell_1^-)\times\cA_3(\widehat{k+1}^+,\ell_2^-,\widehat{Q}^{Sc})~,~~~
\label{partAmm}
\eea

We write them explicitly
\bea
PartA(\widehat{mQ}^-,\ell_1^-)\times\cA_3(\widehat{k+1}^+,\ell_2^-,\widehat{Q}^+)
&=&\spaa{\widehat{mQ}^\bot~\ell_1^\bot}^2\spbb{\widehat{mQ}^\bot~\widehat{k+1}}\over
\spbb{\widehat{k+1}~\ell_2^\bot}\spbb{\ell_2^\bot~\widehat{mQ}^\bot}~~~~
\nonumber\\
&=&{{-}\spab{\ell_1^\bot|\widehat{mQ}^\bot|q}^2\spab{q|\widehat{mQ}^\bot|\widehat{k+1}}^3\over
\spbb{\widehat{k+1}~\ell_2^\bot}\spab{q|\widehat{mQ}^\bot|q}^2\spab{q|\widehat{mQ}^\bot|\ell_2^\bot}}~,~~~\nn
PartA(\widehat{mQ}^+,\ell_1^-)\times\cA_3(\widehat{k+1}^+,\ell_2^-,\widehat{Q}^-)
&=&-\mu^2\spaa{\ell_1^\bot~q}^2\spaa{\ell_2^\bot~\widehat{mQ}^\bot}^3\over
\spaa{\widehat{k+1}~\widehat{mQ}^\bot}\spaa{\widehat{k+1}~\ell_2\bot}\spaa{q~\widehat{mQ}^\bot}^2~~~~
\nonumber\\
&=&{-\mu^2\spaa{\ell_1^\bot~q}^2\spab{\ell_2^\bot|\widehat{mQ}^\bot|q}^3\over
\spaa{\widehat{k+1}~\widehat{mQ}^\bot}\spab{q|\widehat{mQ}^\bot|q}^2\spab{\widehat{k+1}|\widehat{mQ}^\bot|q}}~,~~~\nn
\sum_{Sc} PartA(\widehat{mQ}^{Sc},\ell_1^-)\times\cA_3(\widehat{k+1}^+,\ell_2^-,\widehat{Q}^{Sc})
&=&2\mu^2\spaa{\widehat{mQ}^\bot~\ell_1^\bot}\spaa{\ell_1^\bot~q}\spaa{\ell_2^\bot~q}\spaa{\ell_2^\bot~\widehat{mQ}^\bot}^2\over
\spaa{\widehat{k+1}~\ell_2^\bot}\spaa{\widehat{k+1}~\widehat{mQ}^\bot}\spaa{\widehat{mQ}^\bot~q}^2~~~~
\nonumber\\
&=&{2\mu^2\spaa{\ell_1^\bot~q}\spaa{\ell_2^\bot~q}\spab{\ell_1^\bot|\widehat{mQ}^\bot|q}\spab{\ell_2^\bot|\widehat{mQ}^\bot|q}^2\over
\spaa{\widehat{k+1}~\ell_2^\bot}\spab{q|\widehat{mQ}^\bot|q}^2\spab{\widehat{k+1}|\widehat{mQ}^\bot|q}}~,~~~
\label{mm}
\eea

So
\bea
&&\sum_{h} PartA(\widehat{mQ}^h,\ell_1^-)\times A_R(\widehat{k+1}^+,\ell_2^-,m\WH{Q}^{-h})
\nonumber\\
&=& {{-\spab{\ell_1^\bot|\widehat{mQ}^\bot|q}^2\spab{q|\widehat{mQ}^\bot|\widehat{k+1}}^3}\over
\spbb{\widehat{k+1}~\ell_2^\bot}\spab{q|\widehat{mQ}^\bot|q}^2\spab{q|\widehat{mQ}^\bot|\ell_2^\bot}}
+ {-\mu^2\spaa{\ell_1^\bot~q}^2\spab{\ell_2^\bot|\widehat{mQ}^\bot|q}^3\over
\spaa{\widehat{k+1}~\widehat{mQ}^\bot}\spab{q|\widehat{mQ}^\bot|q}^2\spab{\widehat{k+1}|\widehat{mQ}^\bot|q}}~~~~
\nonumber\\
&+& {\mu^2\spaa{\ell_1^\bot~q}\spaa{\ell_2^\bot~q}\spab{\ell_1^\bot|\widehat{mQ}^\bot|q}
\spab{\ell_2^\bot|\widehat{mQ}^\bot|q}^2\over
\spaa{\widehat{k+1}~\ell_2^\bot}\spab{q|\widehat{mQ}^\bot|q}^2\spab{\widehat{k+1}|\widehat{mQ}^\bot|q}}
+{\mu^2\spaa{\ell_1^\bot~q}\spaa{\ell_2^\bot~q}\spab{\ell_1^\bot|\widehat{mQ}^\bot|q}
\spab{\ell_2^\bot|\widehat{mQ}^\bot|q}^2\over
\spaa{\widehat{k+1}~\ell_2^\bot}\spab{q|\widehat{mQ}^\bot|q}^2\spab{\widehat{k+1}|\widehat{mQ}^\bot|q}}~~~~
\nonumber\\
&=&{ {-\mu^2\spaa{\ell_2^\bot~q}\spab{\ell_1^\bot|\widehat{mQ}^\bot|q}^2\spab{\ell_2^\bot|\widehat{mQ}^\bot|q}\over
\spaa{\widehat{k+1}~\ell_2^\bot}\spab{q|\widehat{mQ}^\bot|q}^2\spab{\widehat{k+1}|\widehat{mQ}^\bot|q}}}
+ {{\mu^2\spaa{\ell_1^\bot~\ell_2^\bot}\spaa{\ell_2^\bot~q}\spab{\ell_1^\bot|\widehat{mQ}^\bot|q}
\spab{\ell_2^\bot|\widehat{mQ}^\bot|q}}\over
\spaa{\widehat{k+1}~\ell_2^\bot}\spab{q|\widehat{mQ}^\bot|q}\spab{\widehat{k+1}|\widehat{mQ}^\bot|q}}
\nonumber\\
&+&
{\mu^2\spaa{\ell_1^\bot~\ell_2^\bot}\spaa{\ell_1^\bot~q}\spab{\ell_2^\bot|\widehat{mQ}^\bot|q}^2\over
\spaa{\widehat{k+1}~\ell_2^\bot}\spab{q|\widehat{mQ}^\bot|q}\spab{\widehat{k+1}|\widehat{mQ}^\bot|q}}
+ {\spaa{\ell_2^\bot~q}^3\spab{\ell_1^\bot|\widehat{mQ}^\bot|q}^2\spbb{\ell_2^\bot~{k+1}}^2\over
\spab{q|\widehat{mQ}^\bot|q}^2\spab{q|\widehat{mQ}^\bot|\ell_2^\bot}}~~~~
\nonumber\\
&=&{1\over\spab{q|\widehat{mQ}^\bot|q}^2}({-\mu^2\spaa{\ell_2^\bot~q}\spab{\ell_1^\bot|\widehat{mQ}^\bot|q}^2
\spab{\ell_2^\bot|\widehat{mQ}^\bot|q}\over\spaa{\widehat{k+1}~\ell_2^\bot}\spab{\widehat{k+1}|\widehat{mQ}^\bot|q}}
-{\spaa{\ell_2^\bot~q}^3\spab{\ell_1^\bot|\widehat{mQ}^\bot|q}^2\spbb{\ell_2^\bot~{k+1}}^2
\over\spab{q|\widehat{mQ}^\bot|\ell_2^\bot}})~~~~\
\nonumber\\
&+&{1\over\spab{q|\widehat{mQ}^\bot|q}}({{\mu^2\spaa{\ell_1^\bot~\ell_2^\bot}\spaa{\ell_2^\bot~q}
\spab{\ell_1^\bot|\widehat{mQ}^\bot|q}\spab{\ell_2^\bot|\widehat{mQ}^\bot|q}}
\over\spaa{\widehat{k+1}~\ell_2^\bot}\spab{\widehat{k+1}|\widehat{mQ}^\bot|q}}
+{\mu^2\spaa{\ell_1^\bot~\ell_2^\bot}\spaa{\ell_1^\bot~q}\spab{\ell_2^\bot|\widehat{mQ}^\bot|q}^2
\over\spaa{\widehat{k+1}~\ell_2^\bot}\spab{\widehat{k+1}|\widehat{mQ}^\bot|q}})
\nonumber\\
&=& {-\mu^2\spaa{\ell_1^\bot~\ell_2^\bot}^2\spab{\ell_2^\bot|\widehat{mQ}^\bot|q}\over
\spaa{\widehat{k+1}~\ell_2^\bot}\spab{\widehat{k+1}|\widehat{mQ}^\bot|q}}~~~~
\nonumber\\
&=& {\spaa{\ell_1^\bot~\ell_2^\bot}^2\spab{k|\ell_2|k+1}\over\spaa{k~{k+1}}}~.~~~
\label{reducemm}
\eea

For the first equality we substitute (\ref{mm}) into (\ref{partAmm}). To get the second equality, we technically use the Schouten Identity as (\ref{schouten})
to factorize the last two terms
\bea
\spaa{\ell_1^\bot~q}\spab{\ell_2^\bot|\widehat{mQ}^\bot|q}
&=&\spaa{\ell_1^\bot~\ell_2^\bot}\spab{q|\widehat{mQ}^\bot|q}
+\spab{\ell_1^\bot|\widehat{mQ}^\bot|q}\spaa{\ell_2^\bot~q}~,~~~\nn
\spaa{\ell_2^\bot~q}\spab{\ell_1^\bot|\widehat{mQ}^\bot|q}
&=&\spaa{\ell_2^\bot~\ell_1^\bot}\spab{q|\widehat{mQ}^\bot|q}
+\spab{\ell_2^\bot|\widehat{mQ}^\bot|q}\spaa{\ell_1^\bot~q}~.~~~
\label{schouten}
\eea
And for the third equality in (\ref{reducemm}), we combine terms according to the power of ${\spab{q|\widehat{mQ}|q}}$. To get the fourth equality, we apply (\ref{schouten}) and (\ref{identities}) technically to eliminate ${\spab{q|\widehat{mQ}|q}}$ and get remaining term without ${\spab{q|\widehat{mQ}|q}}$. The final equality is easy to get using (\ref{identities}).

\section{All-plus 1-loop integrand in $\mathcal{Q}$-cut representation}
\label{qcut}
After we get general expressions for these tree amplitudes, we can use them to calculate the compact loop integrand for color-ordered n-point gluon amplitude $\cA^{\oneloop}_n(1^+,2^+,\dots,n^+)$ by $\mathcal{Q}$-cut construction.

The content below follows the calculation for $A_4^{\oneloop}(1^+,2^+,3^+,4^+)$ in \cite{Huang:2015cwh}.
In the following steps,  we find 
that the product $\cA_L\cA_R$ are independent of $q$, thus the loop integrand is also independent of
$q$, which serves as a consistency check for our tree amplitudes.
 
The ${\cal Q}$-cut
representation
 of the loop integrand is
\bea \cI_{n}^{\cal Q}(\ell)=\sum_{P_L}\sum_{h_1,h_2}\cA_{L}(\cdots,
\widehat{\ell}_R^{~h_1},-\widehat{\ell}_L^{~h_2}){1\over
\W\ell^2}{1\over (-2\W\ell\cdot
P_L+P_L^2)}\cA_R(\widehat{\ell}_L^{~\bar{h}_2},-\widehat{\ell}_R^{~\bar{h}_1},\cdots)~,
~~~\label{q-cut}\eea
where $\widehat{\ell}=\alpha_L(\ell+\eta)$, $\widehat{\ell}_R\equiv
\widehat{\ell}_L-P_L$ with $\alpha_L=P_L^2/(2\ell\cdot P_L)\neq 0$,
$\eta^2=\ell^2$. We review shortly that two deformations have been applied to the loop
momentum $\ell$: the first one is the dimensional deformation $\ell \to
\ell+\eta$ with $\eta$ in extra dimensions, and the second one is the scale deformation $\ell\to \alpha\ell$.
The details of the one-loop $\mathcal{Q}$-cut construction was clarified in \cite{Huang:2015cwh}, and generalizations to two
loops or more was illustrated in \cite{Baadsgaard:2015twa}.

According to (\ref{q-cut}), the ${\cal Q}$-cut
representation of the n-point all-plus 1-loop integrand is given by
\bea
\cI^{\mathcal{Q}}(\widetilde{\ell})
&=&
\sum_{k=2}^{n-2} \sum_{h_1,h_2}
\cA_{L}(1^+,2^+,\dots,k^+,\widehat{\ell}_R^{~h_1},-\widehat{\ell}_L^{~h_2})
{1\over
\widetilde{\ell}^2}{1\over (-2\widetilde{\ell}\cdot
p_{L}+p_{L}^2)}
\cA_R(\widehat{\ell}_L^{~\bar{h}_2},-\widehat{\ell}_R^{~\bar{h}_1},(k+1)^+,\cdots,n^+)
\nonumber\\
&&+\cyclic{p_1,p_2,\dots,p_{n-1},p_n}
\nonumber\\
&=&\sum_{k=2}^{n-2}  (
\cA_{L}(1^+,2^+,\dots,k^+,\widehat{\ell}_R^{~+},-\widehat{\ell}_L^{~-})
\cA_R(\widehat{\ell}_L^{~-},-\widehat{\ell}_R^{~+},(k+1)^+,\cdots,n^+)
\nonumber\\
&&+
\cA_{L}(1^+,2^+,\dots,k^+,\widehat{\ell}_R^{~-},-\widehat{\ell}_L^{~+})
\cA_R(\widehat{\ell}_L^{~+},-\widehat{\ell}_R^{~-},(k+1)^+,\cdots,n^+)
\nonumber\\
&&+
\sum_{S_A}\cA_{L}(1^+,2^+,\dots,k^+,\widehat{\ell}_R^{~S_A},-\widehat{\ell}_L^{~S_A})
\cA_R(\widehat{\ell}_L^{~S_A},-\widehat{\ell}_R^{~S_A},(k+1)^+,\cdots,n^+)~){1\over
\widetilde{\ell}^2}{1\over (-2\widetilde{\ell}\cdot
p_{L}+p_{L}^2)}
\nonumber\\
&+&\cyclic{p_1,p_2,\dots,p_{n-1},p_n}
~.~~~\eea

Writing the $D$-dimensional vector as
$\widehat{\ell}=(\ell,\mu,\eta)$, we have
\bea 
&&\cA_{L}(1^+,2^+,\dots,k^+,
\widehat{\ell}_R^{~+},-\widehat{\ell}_L^{~-})\cA_R(\widehat{\ell}_L^{~+},-\widehat{\ell}_R^{~-},(k+1)^+,\dots,n^+)
\nonumber\\
&=&
{{(\mu^2+\eta^2)^2}
\over \spaa{1~2}\cdots\spaa{k-1~k} \spaa{k+1~k+2}\cdots \spaa{n-1~n}}
\nonumber\\
&\times&
{\spbb{1|\prod_{i=2}^{k-1}(\mu^2+\eta^2-x_{i,\ell_L}x_{\ell_L,i+1})|k}\spbb{k+1|\prod_{i=k+2}^{n-1}(\mu^2+\eta^2-x_{i,\ell_R}x_{\ell_R,i+1})|n}
\over\prod_{i=2}^{k}(\mu^2+\eta^2+x_{\ell_L,i}^2)\prod_{i=k+2}^{n}(\mu^2+\eta^2+x_{\ell_R,i}^2)}~.~~~
\eea

Writing the $D$-dimensional vector as
$\widehat{\ell}=(\ell,\mu,\eta)$, we have
\bea 
&&\cA_{L}(1^+,2^+,\dots,k^+,
\widehat{\ell}_R^{~+},-\widehat{\ell}_L^{~-})\cA_R(\widehat{\ell}_L^{~+},-\widehat{\ell}_R^{~-},(k+1)^+,\dots,n^+)
\nonumber\\
&=&
{{(\mu^2+\eta^2)^2}
\over \spaa{1~2}\cdots\spaa{k-1~k} \spaa{k+1~k+2}\cdots \spaa{n-1~n}}
\nonumber\\
&\times&
{\spbb{1|\prod_{i=2}^{k-1}(\mu^2+\eta^2-x_{i,\ell_L}x_{\ell_L,i+1})|k}\spbb{k+1|\prod_{i=k+2}^{n-1}(\mu^2+\eta^2-x_{i,\ell_R}x_{\ell_R,i+1})|n}
\over\prod_{i=2}^{k}(\mu^2+\eta^2+x_{\ell_L,i}^2)\prod_{i=k+2}^{n}(\mu^2+\eta^2+x_{\ell_R,i}^2)}~.~~~
\eea

The rest two diagrams that contribute are $$\cA_{L}(\widehat{\ell}_R^{~+},-\widehat{\ell}_L^{~-})\cA_R(\widehat{\ell}_L^{~+},-\widehat{\ell}_R^{~-})~~\mbox{and}~~\cA_{L}(
\widehat{\ell}_R^{~S_A},-\widehat{\ell}_L^{~S_A})\cA_R(\widehat{\ell}_L^{~S_A},-\widehat{\ell}_R^{~S_A})$$
which leads to exactly the same results. 

Under the massless conditions of $\WH \ell_L, \WH\ell_R$, we can make
the following replacement $\eta^2\to \widetilde{\ell}^2$ where
$\W\ell=(\ell,\mu)$, $\widetilde{\ell}\to
\alpha_L\widetilde{\ell}$ and $\eta\to\alpha_L\eta$, as well as
$\alpha_L=p_{L}^2/(2\widetilde{\ell}\cdot p_{L})$ in succession. After changing $\eta$, the general integrand in
$\mathcal{Q}$-cut representation is
\bea
\cI^{\mathcal{Q}}(\widetilde{\ell})&=&(2-2\epsilon)
\sum_{k=2}^{n-2}{1\over \spaa{1~2}\cdots \spaa{k-1~k} \spaa{k+1~k+2} \cdots \spaa{n-1~n}}
{(\mu^2+\widetilde{\ell}^2)^2(p_{L}^2/(2\widetilde{\ell}\cdot
p_{L}))^2\over \widetilde{\ell}^2(-2\widetilde{\ell}\cdot
p_{L}+p_{L}^2)
}
\nonumber\\
&\times&
{\spbb{1|\prod_{i=2}^{k-1}(\mu^2+\widetilde{\ell}^2-x_{i,\ell_L}x_{\ell_L,i+1})|k}\spbb{k+1|\prod_{i=k+2}^{n-1}(\mu^2+\widetilde{\ell}^2-x_{i,\ell_R}x_{\ell_R,i+1})|n}
\over\prod_{i=2}^{k}(\mu^2+\widetilde{\ell}^2+x_{\ell_L,i}^2)\prod_{i=k+2}^{n}(\mu^2+\widetilde{\ell}^2+x_{\ell_R,i}^2)}
\nonumber\\
&+&
\cyclic{p_1,p_2,\cdots,p_{n-1},p_n}~,~~~\label{YM-4posi-Q} \eea
where we have summed over helicity states in
$(4-2\eps)$-dimension (especially including the $S_A$ components in
$\dim[\mu]=(-2\epsilon)$-dimension).
 
\section{Summary}
Our results for the series of color-ordered tree amplitudes are very concise. Each of these amplitudes shares a common structure where one part of the amplitude relies on helicity difference of the pair of legs with extra-dimensional momenta and the other part containing pole structures is the same for each case.
Our strategies to use the BCFW recursion relations are quite efficient to prove the general expressions for these amplitudes and reveal how different parts of the amplitudes evolve during the recursion.

Conversely, we emphasize that our results are also correct for the associated massive amplitudes,  which can be used in the unitarity cut method for 1-loop amplitudes and to build superamplitudes on the Coulomb branch. Two of our results $\{\ell_2^-,\ell_1^{-}\}$ and $\{\ell_2^-,\ell_1^{S_a}\}$ are completely new. 

Using these tree amplitudes we successfully form the complete 1-loop all-plus integrand with any numbers of gluons, regardless of traditional integral reduction method. This complete integrand also has a good structure despite the fact that it consists of cut constructive parts as well as rational parts, showing the power of the $\mathcal{Q}$-cut construction.

\section*{Acknowledgments}

We would like to thank Bo Feng who led us to this subject and gave us many instructions for this paper. We are grateful to Gang Yang and Song He who gave us inspiring discussions in the Institute of Theoretical Physics, CAS in Beijing. We are  also grateful that Rijun Huang, Qingjun Jin, Kang Zhou and Junjie Rao gave us many technical suggestions. We would like to thank Rutger Boels for showing early results of \cite{Ferrario:2006np}. Yang An wants to show special gratefulness to his family, friends, Zhen Yan, his tutors Mingxing Luo and Xin Wan for encouragement over the past years.  This work is supported by the National Natural
Science Foundation of China (NSFC) with Grant No.11135006, No.11125523 and
No.11575156. 

\bibliographystyle{JHEP}
\bibliography{Extradimamp}

\end{document}